\documentclass[12pt]{article}
\usepackage{amsthm,amssymb,amsmath}

\newtheorem*{Th}{Theorem}
\newtheorem*{pro}{Proposition}

\newtheorem*{lem}{Lemma}

\newcommand{\bX}{\boldsymbol X}

\newcommand{\bx}{\boldsymbol x}

\newcommand{\be}{\boldsymbol e}

\newcommand{\bb}{{\boldsymbol b}}
\newcommand{\bn}{{\boldsymbol n}}
\newcommand{\bv}{{\boldsymbol v}}
\newcommand{\bomega}{{\boldsymbol \xi}}
\newcommand{\bOmega}{{\boldsymbol \Omega}}
\newcommand{\bzeta}{{\boldsymbol \zeta}}

\newcommand{\R}{{\mathbb R}}
\newcommand{\Z}{{\mathbb Z}}

\newcommand{\D}{{\partial}}

\begin{document}

\title{
Discrete Levy Transformations and \\
Casorati Determinant Solutions \\
of  Quadrilateral Lattices}

\author{Q. P. Liu\thanks{On leave of absence from
Beijing Graduate School, CUMT, Beijing 100083, China}
\thanks{Supported by {\em Beca para estancias temporales
de doctores y tecn\'ologos extranjeros en
Espa\~na: SB95-A01722297}}
   $\,$ and Manuel Ma\~nas\thanks{Partially supported by CICYT:
 proyecto PB95--0401}\\
 Departamento de F\'\i sica Te\'orica,
\\ Universidad Complutense,\\
E28040-Madrid, Spain.}

\date{}

\maketitle

\begin{abstract}
Sequences of discrete Levy and 
adjoint Levy transformations for the
multidimensional quadrilateral lattices are studied. 
After a suitable number of iterations we show how
all the relevant geometrical features of the transformed
quadrilateral lattice can be expressed in terms of 
multi-Casorati determinants.
As an example we dress the Cartesian lattice.
\end{abstract}
\newpage


{\bf 1}.
Recently it has been shown that multidimensional quadrilateral 
lattices
are integrable \cite{DQL1} and a number of results about
this system have been
obtained. Let us mention the reduction mechanism based
on the $\bar\D$ formalism \cite{manakov},
the multidimensional circular lattice \cite{cds}, the relation with 
multi-component
KP through the Miwa transformation and geometrical meaning of the 
$\tau$-function
\cite{dmmms} and Darboux and more general transformations 
\cite{mds,dsm}.

Quadrilateral lattice equations,  as a discrete integrable system,
appeared for the
first time in \cite{bk}, however no geometrical understanding 
of this system can be found
there. We should mention that the quadrilateral lattice has
a continuum limit to conjugate nets \cite{Darboux, eisenhart1}. Since
last century \cite{levy} it has been known that there is a 
transformation,
called Levy transformation, that preserves the conjugacy character of 
the net.
This transformation was iterated, for the bidimensional case, 
in \cite{hammond};
and recently \cite{lm} we have used standard techniques in Soliton 
Theory
to obtain closed formulae, in terms of multi-Wro\'nski determinants,
for all the relevant geometrical objects. The analog of this 
transformation,
say discrete Levy, can be found, for the points of the lattice, in 
\cite{bk}
and a detailed geometrical exposition of it
is contained in \cite{dsm}.

The aim of this paper is to obtain similar results
for the quadrilateral lattice as we did with conjugate nets, namely
to iterate the discrete Levy transformation and its adjoint 
\cite{dsm} to get
closed formulae in terms of multi-Casorati determinants for all the 
geometrical
features of the transformed lattice. We should mention that in 
\cite{mds}
it was obtained, for zero background, multi-Casorati determinant 
representations for quadrilateral
lattices; however, the expressions for the tangent vectors and Lam\'e 
coefficients
are much more involved than here and no closed expression is given 
for the 
points of the lattice. Notice that for the Hirota equation (discrete 
KP)
Casorati determinant representations can be found in \cite{o}.

The layout of this letter is as follows. In the first section we 
remind the reader some
basic facts of the quadrilateral lattices and the discrete Levy 
tarnsformation.
Next, in \S 3 we give the main result of this letter that is extended 
to the
adjoint case in \S 4. In \S 5 we briefly indicate how to write
our formulae, expressions depending on discrete multi-Wro\'nski 
determinants,
in terms of multi-Casorati determinants. Finally, in \S 6 we analyze 
a simple
example by applying our main result    to the Cartesian lattice 
\cite{mds}.
We conlude the letter with an Appendix containg the proof of a lemma
in \S 3.

{\bf 2}. A  Multidimensional Quadrilateral Lattice (MQL) \cite{DQL1} 
is
an $N$-dimensional ($N\geq 2$) lattice:
\[
\bx : {\Z}^{N} \rightarrow
{\R}^{D}, \; M\geq N,\quad\bn:=(n_1,\dots,n_N)\in\Z^N, \quad D\geq N
\]
such that each elementary quadrilateral of it is planar.
It can be shown that this condition can be rewritten
as the following discrete  Laplace equation
\begin{equation}\label{laplace}
\Delta_i\Delta_j\bx=(T_{i} A_{ij})\Delta_i\bx+
(T_j A_{ji})\Delta_j\bx \; \; ,\;\; i\not= j, \; \; \;  i,j=1 ,\dots, 
N,
\end{equation}
where $\bx\in\R^D$ is an arbitrary point of the lattice, $A_{ij}$ are
$N(N-1)$ real functions of $\bn$, $T_j$ is the translation operator 
in
the $j$-th variable:
\[
T_j(f(n_1,\dots,n_j,\dots,n_N))=f(n_1,\dots,n_j+1,\dots,n_N)
\]
and $\Delta_j=T_j-1$ is the corresponding difference operator.

The following nonlinear constraints must hold as compatibility
conditions in order to have planarity in each pair of directions:
\begin{equation}\label{aij}
\Delta_k A_{ij} =
 (T_jA_{jk})A_{ij} +(T_k A_{kj})A_{ik} - (T_kA_{ij})A_{ik},
\;\; i\neq j\neq k\neq i, \end{equation} which characterize 
completely all the
 MQL's.

Equations (\ref{laplace}) can be written as first order
systems \cite{DQL1}, for this we introduce  functions $H_i$,
$i=1,\dots,N$, defined by
\begin{equation} \label{gamma}
 \Delta_jH_i=A_{ij}H_i.
\end{equation}
Then (\ref{laplace}) reads
\begin{equation}\label{darbouxx}
\Delta_j \bX_i=(T_jQ_{ij})\bX_j, \; \; \; i\neq j,
\end{equation}
where the scalar functions $Q_{ij}$ and the $D$-dimensional vectors
$\bX_{i}$ are defined by the equations
\begin{align}
\Delta_iH_j & = Q_{ij} T_iH_i, \; \; i\not=j,
\label{defbeta} \\
\Delta_i\bx & = (T_iH_i) \bX_i,  \label{defx}
\end{align}
whose compatibility gives the equations
\begin{equation}\label{disdar}
\Delta_jQ_{ik}=(T_jQ_{ijk})Q_{jk}, \;\; i\neq j\neq k\neq i.
\end{equation}
 Equations (\ref{aij}) (or (\ref{disdar}))
are the multidimensional quadrilateral lattice equations.

Given a solution $\xi_j$ of 
\[
\Delta_k\xi_j=(T_kQ_{jk}) \xi_k,
\]
for each of the $N$ possible directions of the lattice
there is a corresponding discrete Levy transformation
that reads for the $i$-th case:
\begin{align*}
&\bx[1]=\bx-\frac{\Omega(\xi,H)}{\xi_i }{\bX_i},
\\[4mm]
&\begin{cases}
\bX_i[1]=\dfrac{\xi_i\Delta_i\bX_i-
(\Delta_i\xi_i)\bX_i}{\xi_i},\\[3.5mm]
\bX_k[1]=\dfrac{\xi_i\bX_k-\xi_k\bX_i}{\xi_i},
\end{cases}
\\[4mm]
&\begin{cases}
H_i[1]=-\dfrac{\Omega(\xi,H)}{\xi_i},\\
H_k[1]=H_k-Q_{ik}\dfrac{\Omega(\xi,H)}{\xi_i},
\end{cases}
\\[4mm]
&
\begin{cases}
Q_{ik}[1]=-\dfrac{Q_{ik}(\Delta_i \xi_i)-
\xi_i\Delta_i Q_{ik}}{\xi_i},\\[3.5mm]
Q_{ki}[1]=-\dfrac{\xi_k}{\xi_i},\\[3.5mm]
Q_{kl}[1]=-\dfrac{\xi_kQ_{il}-\xi_iQ_{kl}}{\xi_i},
\end{cases}
\end{align*}
where $k,l=1,\dotsc, N$ with $k\neq l \neq i$. Here we have
introduced the potential $\Omega(\xi,H)$ defined by
\[
\Delta_k\Omega(\xi,H)=\xi_kT_kH_k,\quad k=1,\dotsc,N,
\]
which are compatible equations by means of the equations
satisfied by $\xi_k$ and $H_k$.

{\bf 3}. As in the continuum case, i. e. conjugate nets,
we have $N$ different elementary Levy's transformations.
Here we are going to iterate
these transformations to get closed formulae for the 
transformed lattice.
For simplicity we shall assume that in the iteration process 
at least one discrete Levy transformation has been made per
direction in the lattice.

To present our main result, we introduce some convenient notations.
Given any set of functions
 $\{\xi^i_j\}_ {\substack{i=1,\dots,M\\ j=1.\dots, N}}$ we denote by
$W_j[n]$ the following discrete Wro\'nski matrix
\[
W_j[n]:= W_j(\xi^1_j, \dotsc, \xi^M_j):=\begin{pmatrix}
\xi^1_j&\xi^2_j&\dots&\xi^M_j\\
\Delta_j\xi^1_{j}&\Delta_j\xi^2_{j}&\dots&\Delta_j\xi^M_{j}\\
\vdots&\vdots& &\vdots\\
\Delta^{n-1}_j\xi^1_{j}&\Delta^{n-1}\xi^2_{j}&\dots&\Delta^{n-1}_j\xi
^M_j\\
\end{pmatrix}.
\]
For any partition of $M=m_1+m_2+ \dots + m_N$, we construct
a discrete multi-Wro\'nski matrix
\[
{\cal W}:=\begin{pmatrix}W_1[m_1]\\  \vdots \\
W_N[m_N]\end{pmatrix}.
\]

Before going to our main result we need the following  technical 
lemma,
whose proof is given in the appendix:
\begin{lem}
We have the relations
\begin{align}
\Delta_k{|\cal W|}&=T_k|\tilde{\cal W}|,\label{1}\\
|{\cal W}|&=T_k|\bar {\cal W}|,\label{2} 
\end{align}
where $\tilde{\cal W}$ and
$\bar {\cal W}$ are obtained from $\cal W$ by replacing the last row
of the $k$-th block   by $T_k^{-1}\Delta_k^{m_k}\bomega_k$
and $T_k^{-1}\Delta_k^{m_k-1}\bomega_k$, respectively.
\end{lem}

With this at hand we have the following:
\begin{Th}\label{theorem}
Given $M$ functions $\{\xi_i^j\}_{\substack{i=1,\dotsc,N\\ 
j=1,\dotsc,
 M}}$ and
$\bX_i=(X_i^1,\dotsc,X_i^D)^{\operatorname{t}}$, $i=1,\dotsc,N$, all 
of them
solutions of \eqref{darbouxx} and $H_i$, $i=1,\dotsc,N$, solutions
of \eqref{defbeta},
for  given $Q_{ij}$,  then new solutions
$\bX_i[M],H_i[M]$ and $Q_{ij}[M]$ are defined by:
\[
X_i^\ell[M]=\frac{\left|{\mathbb X}_i^\ell\right|}{\left|{\cal
W}\right|},\quad H_i[M]=-\frac{\left|{\mathbb
H}_i\right|}{\left|{\cal W}\right|},\quad
Q_{ij}[M]=-\frac{\left|{\cal W}_{ij}\right|}{\left|{\cal W}\right|},
\]
where
\[
{\mathbb X}_i^\ell=\begin{pmatrix}{\cal W}&\bv^\ell\\
\Delta_i^{m_i}\bomega_i&\Delta_i^{m_i}X_i^\ell\end{pmatrix},
\]
with
\begin{align*}
\bv^\ell&:=(\bv_1^\ell,\dotsc,\bv_N^\ell)^{\operatorname{t}},\text{ 
being }
\bv_k^\ell:=(X_k^\ell,\Delta_k X_k^\ell,\dotsc,
\Delta_k^{m_k-1}X_k^\ell),\\
\bomega_i&:=(\xi_i^1,\dotsc,\xi_i^M),
\end{align*}
${\mathbb H}_i$ is obtained from $\cal W$ by replacing the last row
of the $i$-th block by $\Omega(\bomega,H)$ and ${\cal W}_{ij}$
by replacing the last row of the $j$-th block by  
$\Delta_i^{m_i}\bomega_i$.
In the partition $M=m_1+m_2+ \dots + m_N$ we need $m_i\in\mathbb N$.

Moreover, for the new quadrilateral 
lattice we have the parametrization
\[
\bx[M]=\frac{1}{|{\cal W}|}\bigg(\begin{vmatrix}{\cal W} &\bv^1\\
\Omega(\bomega,H) &x^1
\end{vmatrix},\dotsc,\begin{vmatrix}{\cal W} &\bv^D\\
\Omega(\bomega,H) &x^D
\end{vmatrix} \bigg)^{\operatorname{t}}.
\]
\end{Th}

\begin{proof} 

We first need to show that
\[
\Delta_kX_i^\ell[M]=(T_kQ_{ik}[M])X_k^\ell[M],
\]
or equivalently that the following bilinear equation holds
\[
\left|{\cal W}\right|\Delta_k\left|{\mathbb X}_i^\ell\right|-
\left|{\mathbb X}_i^\ell\right| \Delta_k\left|{\cal W}\right|+
\left|{\mathbb X}_k^\ell\right|T_k\left|{\cal W}_{ik}\right|=0.
\]
To this aim, using standard techniques \cite{freeman}, we consider 
the following $(2M+1)\times(2M+1)$ square
matrix
\[
{\cal A}_{ik}^\ell:=
\begin{pmatrix}
\quad A_k\quad& 0 &\Delta_k^{m_k-1}\bomega_k^{\operatorname{t}}&
\Delta_k^{m_k}\bomega_k^{\operatorname{t}}&
T_k\Delta_i^{m_i}\bomega_i^{\operatorname{t}}\\[3mm]
\quad 0\quad &\quad 
A_k\quad&\Delta_k^{m_k-1}\bomega_k^{\operatorname{t}}&
\Delta_k^{m_k}\bomega_k^{\operatorname{t}}&
T_k\Delta_i^{m_i}\bomega_i^{\operatorname{t}}\\[3mm]
0&\bb_k^\ell&\Delta_k^{m_k-1}X_k^\ell&\Delta_k^{m_k}X_k^\ell&
T_k\Delta_i^{m_i}X_i^\ell
\end{pmatrix},
\]
where $A_k$ is a $M\times(M-1)$ rectangular matrix
\[
(A_k)^{\operatorname{t}}:=T_k
\begin{pmatrix}
 W_1[m_1]\\
\vdots\\
\hat W_k[m_k]
\\
\vdots\\
W_N[m_N]
\end{pmatrix},
\]
with $\hat W_k[m_k]$ obtained from $W_k[m_k]$ by deleting the last 
row,
and
\[
\bb_k^\ell=(\bv_1^\ell,\cdots,\hat\bv_k^\ell,\dotsc,\bv_N^\ell),
\]
with $\hat\bv_k^\ell$ obtained by deleting the last element in 
$\bv_k^\ell$.
Applying  our lemma and the  Laplace's general expansion theorem 
\cite{algebra} to compute $\det{\cal A}_{ik}^\ell$,
 we obtain the desired bilinear relation.

Next, let us check the relation
\[
\Delta_kH_i[M]=Q_{ki}[M]T_kH_k[M],
\]
or equivalently that the following bilinear equation holds:
\[
{\left|\cal W\right|}\Delta_k\left|{\mathbb H}_i\right|-
\left|{\mathbb H}_i\right|\Delta_k\left|{\cal W}\right|
+\left|{\cal W}_{ki}\right|T_k\left|{\mathbb H}_k\right|=0.
\]
This formula, as previously, follows from the Laplace's general
expansion theorem when applied to the evaluation of the
determinant of the  $2M\times 2M$ matrix:
\[
{\cal B}_{ik}:=
\begin{pmatrix}
\quad B_{ik}\quad& 0  &
\Delta_k^{m_k-1}\bomega_k^{\operatorname{t}}&
\Delta_k^{m_k}\bomega_k^{\operatorname{t}}&
T_k\Delta_i^{m_i-1}\bomega_i^{\operatorname{t}}&
T_k\Omega(\bomega,H)^{\operatorname{t}}\\[3mm]
\quad 0\quad &\quad B_{ik}\quad
&\Delta_k^{m_k-1}\bomega_k^{\operatorname{t}}&
\Delta_k^{m_k}\bomega_k^{\operatorname{t}}&
T_k\Delta_i^{m_i-1}\bomega_i^{\operatorname{t}}&
T_k\Omega(\bomega,H)^{\operatorname{t}}
\end{pmatrix},
\]
where $B_{ik}$ is a $M\times(M-2)$ rectangular matrix
\[
(B_{ik})^{\operatorname{t}}:=
T_k\begin{pmatrix}
 W_1[m_1]\\
\vdots\\
\hat W_i[m_i]\\
\vdots\\
\hat W_k[m_k]
\\
\vdots\\
W_N[m_N]
\end{pmatrix}.
\]

Finally, we prove the formula for the points in the 
lattice $x^\ell[M]=\Omega(X^\ell[M],H[M])$ 
(see \eqref{defx}).
 For that aim we consider the following $(2M+1)\times(2M+1)$ matrix
\[
{\cal C}_k^\ell:=
\begin{pmatrix}
\quad A_k\quad& 0 &\Delta_k^{m_k-1}\bomega_k^{\operatorname{t}}&
\Delta_k^{m_k}\bomega_k^{\operatorname{t}}&
T_k\Omega(\bomega,H)^{\operatorname{t}}\\[3mm]
\quad 0\quad &\quad 
A_k\quad&\Delta_k^{m_k-1}\bomega_k^{\operatorname{t}}&
\Delta_k^{m_k}\bomega_k^{\operatorname{t}}&
T_k\Omega(\bomega,H)^{\operatorname{t}}\\[3mm]
0&\bb_k^\ell&\Delta_k^{m_k-1}X_k^\ell&\Delta_k^{m_k}X_k^\ell&T_k
\Omega(X^\ell,H)
\end{pmatrix},
\]
and use  that $x^\ell=\Omega(X^\ell,H)$ 
and compute the $\det{\cal C}^\ell_k$ according to the
Laplace's general expansion theorem.
\end{proof}

{\bf 4}.   The discrete adjoint Levy transformation \cite{dsm}:
\begin{align*}
&\bx[1]=\bx-\frac{\Omega(\bX,\zeta)}{\zeta_i }{H_i},
\\[4mm]
&\begin{cases}
\bX_i[1]=-\dfrac{\Omega(\bX,\zeta)}{\zeta_i},\\
\bX_k[1]=\bX_k-Q_{ki}\dfrac{\Omega(\bX,\zeta)}{\zeta_i},
\end{cases}
\\[4mm]
&\begin{cases}
H_i[1]=\dfrac{\zeta_i\Delta_iH_i-
(\Delta_i\zeta_i)H_i}{\zeta_i},\\[3.5mm]
H_k[1]=\dfrac{\zeta_iH_k-\zeta_kH_i}{\zeta_i},
\end{cases}
\\[4mm]
&
\begin{cases}
Q_{ki}[1]=-\dfrac{Q_{ki}\Delta_i \zeta_i-
(\Delta_iQ_{ki})\zeta_i}{\zeta_i},\\[3.5mm]
Q_{ik}[1]=-\dfrac{\zeta_k}{\zeta_i},\\[3.5mm]
Q_{lk}[1]=-\dfrac{\zeta_kQ_{li}-\zeta_iQ_{lk}}{\zeta_i},
\end{cases}
\end{align*}
where $k,l=1,\dotsc, N$ with $k\neq l \neq i$, and $\zeta_k$ solves
\eqref{defbeta}.

By similar considerations as in \S {\bf 3} we get:
\begin{pro}
Given $M$ functions $\{\zeta_i^j\}_{\substack{i=1,\dotsc,N\\ 
j=1,\dotsc,
 M}}$ and $H_i$, $i=1,\dotsc,N$, , all of them
solutions
of \eqref{defbeta},
$\bX_i=(X_i^1,\dotsc,X_i^D)^{\operatorname{t}}$, $i=1,\dotsc,N$ 
solutions
 of \eqref{darbouxx} for  given $Q_{ij}$,  then new solutions
$\bX_i[M],H_i[M]$ and $Q_{ij}[M]$ are defined by:
\[
X_i^\ell[M]=-\frac{\big|\tilde{\mathbb X}_i^\ell\big|}{\big|{\cal
W}\big|},\quad H_i[M]=\frac{\big|\tilde{\mathbb
H}_i\big|}{\left|{\cal W}\right|},\quad
Q_{ij}[M]=-\frac{\left|{\cal W}_{ji}\right|}{\left|{\cal W}\right|},
\]
where
\[
\tilde{\mathbb H}_i=\begin{pmatrix}{\cal W}&\tilde{\bv}\\
\Delta_i^{m_i}\bzeta_i&\Delta_i^{m_i}H_i\end{pmatrix},
\]
with
\begin{align*}
\tilde{\bv}&:=(\tilde{\bv}_1,\dotsc,\tilde{\bv}_N)^{\operatorname{t}}
,\text{ being }
\tilde{\bv}_k:=(H_k,\Delta_k H_k,\dotsc,
\Delta_k^{m_k-1}H_k),\\
\bzeta_i&:=(\zeta_i^1,\dotsc,\zeta_i^M),
\end{align*}
${\mathbb X}_i^\ell$ is obtained from $\cal W$ by replacing the last 
row
of the $i$-th block by $\Omega(X^\ell,\bzeta)$.
In the partition $M=m_1+m_2+ \dots + m_N$ we need $m_i\in\mathbb N$.

For the new quadrilateral lattice we have the parametrization
\[
\bx[M]=\frac{1}{|{\cal W}|}\bigg(\begin{vmatrix}{\cal W} 
&\tilde{\bv}^1\\
\Omega(X^1,\bomega) &x^1
\end{vmatrix},\dotsc,\begin{vmatrix}{\cal W} &\tilde{\bv}^D\\
\Omega(X^P,\bomega) &x^D
\end{vmatrix} \bigg)^{\operatorname{t}}.
\]
\end{pro}

{\bf 5}. The result of our Theorem and Proposition are expressed in 
terms
of discrete multi-Wro\'nski matrices, however it is easy to write the 
formulae in 
terms of  multi-Casorati determinants. For this aim it is only 
necessary to replace
in ${\cal W},\mathbb X_i^\ell,\tilde\mathbb H_i,\bv,\tilde\bv$
and ${\cal W}_{ij}$ the difference operator $\Delta$ by the shift 
operator
$T$.

{\bf 6}. We now consider the dressing of the Cartesian background 
\cite{mds}.
When $Q_{ij}=0$ the quantities $\bX_i$ and $H_i$ are arbitrary 
vector  functions of $n_i$ only, while the points in the
lattice can be represented as 
\[
\bx = \sum_{k=1}^{D}\bx_k(n_k)  + {\boldsymbol c},
\]
where $\bx_k(0)={\boldsymbol 0}$ and ${\boldsymbol c}$ is a constant 
vector,
which characterize this parallelogram lattice \cite{mds}. Among them,
the Cartesian lattice is obtained
choosing $D=N$ and
 \begin{align*}
 &\bX_i=\be_i, \;  H_i =1,\\
 &\bx=\bn,
\end{align*}
where $\{\be_i\}_{i=1}^N$ is the canonical basis of $\R^N$.

Next, in order to get the iterated Levy transformed lattice
we apply our Theorem to these data. First, 
notice that $\bomega_i(\bn)=\bomega_i(n_i)$,
$i=1,\dotsc, N$. Second, we have 
$\bv^\ell_k=(\delta_k^\ell,0,\dotsc,0)$,
where $\delta^\ell_k$ is the Kronecker symbol.
Observe also that $\bOmega:=\Omega(\bomega,H)$ is defined by 
$\Delta_i\bOmega=\bomega_i$, and we 
can write $\bOmega(\bn)=\bOmega_1(n_1)+\dotsb+\bOmega_N(n_N)$ and
$\bomega_i=\Delta_i\bOmega_i$.
 Thus, we have that
$\mathbb X^\ell_i$ is just the matrix obtained from $\cal W$ by 
replacing
the first row of the $\ell$-th block, $W_\ell$, by 
$\Delta_i^{m_i}\bomega_i$.
To get $\mathbb H_i$ we  subsistute the last row of the $i$-th block
of $\cal W$ by $-\bOmega$. Finally,  the points of the lattice are
\[
\bx(\bn)=\bn-\delta\bx(\bn),
\quad\delta x_i:=\frac{|{\cal W}_i|}{|{\cal W}|},
\] 
where ${\cal W}_i$ is the matrix obtained
from $\cal W$ by replacing the first row of the $i$-th block, $W_i$, 
by
$\bOmega$.

The result of the iterated adjoint Levy transformation on the 
Cartesian
lattice, as one can easily compute from our Proposition, gives the 
same
transformed lattice as above. 
\\[3mm]

{\bf Appendix:} In this appendix
we give the proof of our lemma. 
\begin{proof}[Proof of the Lemma]
For the sake of simplicity we shift the $k$-th block to the end of 
the matrix
and use the discrete version of the Leibnitz rule,
$\Delta(ab)=(\Delta a)b+(Ta)\delta b$.
Then, we can write
\[
\Delta_k|{\cal W}|={\cal F}_1+\dotsb+{\cal F}_N
\]
with 
\[
{\cal F}_j:={\cal F}_{j1}+\dotsb+{\cal F}_{jm_j},\, j\neq k,
\, {\cal F}_{ji}:=
\begin{vmatrix}
T_kW_1[m_1]\\\vdots\\T_kW_{j-1}[m_{j-1}]\\
\tilde W_{ji}[m_j]\\ W_{j+1}[m_{j+1}]\\\vdots\\W_k[m_k]
\end{vmatrix},\, 
\tilde W_{ji}[m_j]:=
\begin{pmatrix}
T_k\bomega_j\\
T_k\Delta_j\bomega_j\\
\vdots\\
T_k\Delta_j^{i-1}\bomega_j\\
\Delta_k\Delta_j^i\bomega_j\\
\Delta_j^{i+1}\bomega_j\\
\vdots\\
\Delta_j^{m_j-1}\bomega_j
\end{pmatrix}.
\]
We can remove the $T_k$, namely
\[
{\cal F}_{ji}=
\begin{vmatrix}
W_1[m_1]\\\vdots\\W_{j-1}[m_{j-1}]\\
\breve W_{ji}[m_j]\\ W_{j+1}[m_{j+1}]\\\vdots\\W_k[m_k]
\end{vmatrix},\;
\breve W_{ji}[m_j]:=
\begin{pmatrix}
\bomega_j\\
\Delta_j\bomega_j\\
\vdots\\
\Delta_j^{i-1}\bomega_j\\
\Delta_k\Delta_j^i\bomega_j\\
\Delta_j^{i+1}\bomega_j\\
\vdots\\
\Delta_j^{m_j-1}\bomega_j
\end{pmatrix},
\]
because $T_k\Delta_i^p\bomega_i$ can be expressed, using 
\eqref{darbouxx},
as a linear combination:
\[
T_k\Delta_i^p\bomega_i=\Delta_i^p\bomega_i+
c_{k,i,p-1}\Delta_i^{p-1}\bomega_i+\dotsb+
c_{k,i,1}\Delta_i\bomega_i+
c_{k,i,0}\bomega_i+c_{k,i}\bomega_k
\]
for some scalar coefficients $c$'s.
Using this formula again we see that ${\cal F}_{ji}=0$.
So, we have 
\[
\Delta_k|{\cal W}|={\cal F}_k
\]
where
\[
{\cal F}_k:={\cal F}_{k1}+\dotsc+{\cal F}_{km_k},\;
{\cal F}_{kp}:=\begin{vmatrix}
T_kW_1[m_1]\\\vdots\\T_kW_N[m_N]\\\breve W_{kp}[m_k]
\end{vmatrix},\;
\breve W_{kp}[m_k]:=\begin{pmatrix}
T_k\bomega_k\\
T_k\Delta_k\bomega_k\\
\vdots\\
T_k\Delta_k^{p-1}\bomega_k\\
\Delta_k\Delta_k^p\bomega_k\\
\Delta_k^{p+1}\bomega_k\\
\vdots\\
\Delta_p^{m_p-1}\bomega_p
\end{pmatrix}.
\]
Now, it is obvious that 
\[
{\cal F}_k={\cal F}_{km_k},
\]
which gives the \eqref{1}.
For \eqref{2} one proceeds in a similar way.
\end{proof}

\end{document}